# Microscopic optical buffering in a harmonic potential

M. Sumetsky

Aston Institute of Photonic Technologies, Aston University, Birmingham B4 7ET, UK

m.sumetsky@aston.ac.uk

In the early days of quantum mechanics, Schrödinger noticed that oscillations of a wave packet in a one-dimensional harmonic potential well are periodic and, in contrast to those in anharmonic potential wells, do not experience distortion over time. This original idea did not find applications up to now since an exact one-dimensional harmonic resonator does not exist in nature and has not been created artificially. However, an optical pulse propagating in a bottle microresonator (a dielectric cylinder with a nanoscale-high bump of the effective radius) can exactly imitate a quantum wave packet in the harmonic potential. Here, we propose a tuneable microresonator that can trap an optical pulse completely, hold it as long as the material losses permit, and release it without distortion. This result suggests the solution of the long standing problem of creating a microscopic optical buffer, the key element of the future optical signal processing devices.

1. **Introduction**

One of the greatest challenges of the modern photonics is the creation of miniature low-loss and high speed optical signal processors which promise to revolutionize the future computing and communications [1, 2]. The key and most challenging element of these devices enabling controlling and manipulating of optical pulses is the microscopic optical buffer. Having the smallest possible dimensions, the buffer should trap optical pulses, hold them for the required (usually nanosecond-order) period of time, and release them without distortion. The delay of an optical pulse in a small-size photonic structure assumes that the pulse experiences many oscillations (e.g., reflections and rotations) before being released, i.e., its propagation speed averaged over these oscillations is slow. Searching for realistic miniature optical delay lines and buffers based on this "slow light" concept resulted in several remarkable designs employing coupled ring resonators and photonic crystal waveguides [1-6]. The most important of these slow light structures are periodic and their transmission band has a region of approximately zero dispersion, which ensures the nearly dispersionless and slow propagation of optical pulses. These structures though suffer from the bandwidth-delay time limitation, which restricts the values of the delay time and pulse bandwidth achieved simultaneously [1, 2, 7]. A way to overcome this limitation suggested in [6] consists in creation of a miniature optical buffer by adiabatic compression of the transmission band of a coupled resonator optical waveguide. Ideally this device enables slowing down and stopping of a light pulse with the predetermined spectral width. Yet, the experimental realization of all these models encountered significant practical barriers due to insufficient precision of modern photonics technologies and attenuation of light [8, 9]. Consequently, it was suggested that "slow light dispersion, bandwidth, and loss are fundamental issues that will limit the use of slow light devices as buffers" [10].

However, in 1926, soon after the creation of quantum mechanics, Erwin Schrödinger published a paper where he investigated the oscillations of a Gaussian wave packet (used as a model of a quantum particle) in a harmonic potential [11]. He showed that the particle oscillations are periodic, i.e., the wave packet in a harmonic potential does not experience any distortion after many oscillations. Schrödinger wrote: "*Our wave group always remains compact,* and does *not* spread out into larger regions as time goes on, as we were accustomed to make it do, for example, in optics. It is admitted that this does not mean much in one dimension…" While the quantum motion of wave packets in a one-dimensional harmonic potential is indeed difficult to find in nature, the wave packet dynamics in potential wells with more general anharmonic shapes and higher dimensions has been investigated both theoretically and experimentally in atomic and solid state physics. It has been found that, unlike in one-dimensional harmonic potential, this motion is generally not periodic, though often exhibits the revival behaviour [12].

Remarkably, Schrödinger's result can be obtained without solving the Schrödinger equation for the potential with the quadratic dependence on the coordinate. In fact, it is sufficient to assume that the potential well has an equidistance spectrum $E_n = n\omega + E_0$ corresponding to the stationary wave functions $\psi_n(z)$.

Then, from the linearity of quantum mechanics, the evolution of a wave packet $\Psi(z,t)$ with arbitrary initial shape $\Psi(z,0)$ is found as $\Psi(z,t) = \sum_{n=0}^{\infty} a_n \psi_n(z) e^{-in\omega t}$ which has the period $T = 2\pi/\omega$. Thus, the periodicity of wave packet oscillations follows from the equidistance of the spectrum of the resonator within the spectral width of the wave packet, which can be ensured by potentials with more general spatial dependencies.

We note that, in photonics, similar resonators (i.e., those having the equidistance spectrum within the pulse spectral width) can serve as ideal miniature optical buffers since they can hold optical pulses without distortion. In contrast to quantum mechanics, where the experimental realization of such resonators is problematic and still does not mean much in one dimension [11], essentially one-dimensional resonant structures can be realized based on the photonic crystal waveguides, sequences of ring resonators, and fibre Bragg gratings. To this end, the periodicity of these structures should be appropriately chirped [13-15] to arrive at the locally precise equidistant spectrum. Yet, similar to the above mentioned approaches based on the subwavelength-scale modulation of refractive index, the practical realization of these structures is impeded by insufficient fabrication precision and substantial attenuation of light [3, 8-10].

On the other hand, the recently developed photonic fabrication platform, Surface Nanoscale Axial Photonics (SNAP), precisely imitates the one-dimensional Schrödinger equation optically and, at the same time, does not require the subwavelength-scale modulation of the refractive index to arrive at the effective potential with required spectrum governing the slow light propagation [16-18]. The photonic structures in SNAP are created at the surface of an exceptionally smooth and uniform optical fibre by its nanoscale deformation with the unprecedented subangstrom precision. Instead of periodicity, which warrants the slow light propagation in photonic crystals, this platform explores whispering gallery modes, which experience multiple transverse circulations and slowly propagate along the fibre axis. In particular, a SNAP bottle resonator [16-18] with the parabolic effective radius variation can accurately reproduce the Schrödinger's harmonic potential well.

Here, we propose a miniature bottle resonator optical buffer based on the generalised tuneable Schrödinger harmonic oscillator introduced below. This device presents a feasible solution of the long standing problem of creating a smallest possible microscopic buffer for processing of optical pulses. In section 2, to illustrate the idea of the paper, we compare oscillations of an optical pulse launched into the stationary harmonic resonator, which exhibits no distortion, and anharmonic (rectangular) bottle resonator, which exhibits dramatic distortion. In section 3, we construct a non-stationary parabolic potential corresponding to a tuneable bottle resonator and show that it can work as a perfect optical buffer, i.e., it can trap an optical pulse, hold it for a predetermined period of time, and release it without distortion. In section 4, we introduce buffers based on harmonic non-parabolic potentials and show that these buffers can be designed so that the optical pulse is completely undisturbed in the process of switching. Section 5 discusses the fabrication precision of SNAP structures. In section 6, we take into account fabrication errors and show that the proposed miniature optical buffer is feasible. Section 7 discusses and summarises the results obtained.

## 2. Optical pulse oscillations in a potential well

Let us first establish the correspondence between the Schrödinger equation, which describes the motion of a one-dimensional quantum particle [19], and the Schrödinger equation, which describes propagation of light in the SNAP platform. Due to the very small and smooth effective radius variation of a SNAP structure, a whispering gallery mode (WGM) can be determined by separation of variables in cylindrical coordinates $(z, \rho, \varphi)$ as $\exp(im\varphi) \Xi_{mp}(\rho) \Psi_{mp}(z,t)$. Below we consider the resonant propagation of a WGM pulse corresponding to the fixed azimuthal and radial quantum numbers $m$ and $p$, which is fully described by the amplitude $\Psi(z,t)$ as a function of axial coordinate $z$ and time $t$ (here and below the quantum number indexes are omitted for brevity). The equation that determines this propagation has the form of the one-dimensional Schrödinger equation [16], which for the non-stationary case under consideration takes the form:

$$i\mu\frac{\partial \Psi}{\partial t} = -\frac{\partial^2 \Psi}{\partial z^2} + V(z,t)\Psi .\tag{1}$$

Here $\mu = 2\omega_0 n_0^2 / c^2$ and potential $V(z,t) = -2k^2 \Delta r_{eff}(z,t)/r_0$ are defined through the radiation frequency $\omega_0$ of the transmission channel, refractive index $n_0$ and bulk propagation constant $k = \omega_0 n_0 / c$ of the bottle resonator material, speed of light in vacuum $c$, and fibre radius $r_0$. The nanoscale effective variation $\Delta r_{eff}$ of the fibre radius is expressed through the variation of the effective physical radius $\Delta r$ and refractive index $\Delta n$ as $\Delta r_{eff}(z,t) = (\Delta r(z,t)/r_0 + \Delta n(z,t)/n_0)r_0$ [16]. Eq. (1) can be derived from Maxwell equations in a way similar to that used in the derivation of the stationary Schrödinger equation which describes propagation of WGM in a SNAP fibre with nanoscale effective radius variation [16]. However, it is more straightforward to derive this equation directly from the stationary Schrödinger equation by the Fourier transform [20]. In fact, assuming that the potential in Eq. (1) is independent of time, we look for the solution of this equation in the form $\Psi(z,t) = \Phi(z)\exp(i\Delta\omega t)$, where $\Delta\omega$ is the frequency variation. After the substitution of this expression into Eq. (1) we arrive at the known stationary Schrödinger equation for $\Phi(z)$ [16]. The inverse Fourier transform yields Eq. (1). Experimental results [17, 18] demonstrate the fabrication of SNAP structures with sub-angstrom precision in effective radius variation, while the fabrication precision of 0.1 angstrom and better is also feasible.

To illustrate the effect of dispersion and self-interference, we first consider the propagation of an optical pulse launched into the *rectangular* bottle resonator having the height $\Delta r_0 = 2$ nm and length 2 mm (Fig. 1(a)). Here and below, we consider the propagation of 100 ps pulses and set the fibre radius $r_0 = 20$ μm, refractive index $n = 1.5$, and radiation wavelength $\lambda = 1.5$ μm. The spatial width of the 100 ps Gaussian pulse amplitude is 4.2 cm in vacuum and approximately 2.8 cm in a silica fibre. Following the experimental observations [18], we set the initial *axial* speed of the pulse to the realistic 1% of its actual speed in silica, i.e., to $0.0066c$. Consequently, the axial width of the pulse amplitude is reduced to 0.28 mm. The surface plot describing the evolution of this pulse along the fibre axis $z$ as a function of time is shown in Fig. 1(b). It is seen that the pulse experiences significant corruption in the process of bouncing caused by both the dispersion and self-interference [12, 21]. As the result, the original shape of the pulse is completely lost in a few nanoseconds. In contrast, as was first noted by Schrödinger [11], oscillations of an optical pulse in a bottle resonator with quadratic radius variation profile $\Delta r_{eff}(z) = -z^2/(2R)$ and axial radius $R$ are periodic and do not cause dispersion over time as illustrated in Figs. 1(c) and (d).

### 3. Harmonic optical buffering in a harmonic parabolic potential

Generalizing the Schrödinger's result [11], we introduce now a harmonic optical buffer. We show that a tuneable harmonic potential well, which is reproduced by a miniature SNAP bottle resonator illustrated in Fig. 2, can trap an optical pulse completely, hold it as long as the material losses permit, and release without distortion. Light is coupled in and out of this resonator through a transverse waveguide, e.g., a microfibre taper (Fig. 2(a)). The buffering process includes: opening the bottle resonator by nanoscale variation of its effective radius (refractive index) to let the optical pulse in (Fig. 2(b)); closing the resonator when the pulse is completely inside the resonator and holding it for the duration of the required time delay (Fig. 2(c)); and releasing the pulse by reversing the deformation illustrated in Fig. 2(b) (Fig. 2(d)). The pulse dwell time in the optical buffer is not restricted by the delay time-bandwidth limitation since it is determined only by the number of oscillation cycles and material losses which allow the pulse to oscillate in the buffer without significant attenuation. Feasible experimental ways to open and close the bottle resonator by the application of laser and electrical pulses will be described later.

The bandwidth $\Delta\omega$ of the optical pulse that can be held in the bottle resonator is expressed through the magnitude of its effective radius variation $\Delta r_{eff}$ as

$$\frac{\Delta\omega}{\omega_0} \sim \frac{\Delta r_{eff}}{r_0}. \tag{2}$$

For the case of purely refractive index tuning, this equation coincides with $\Delta\omega/\omega_0 \sim \Delta n/n_0$ [6, 22]. The amplitude of the effective radius variation required for opening and closing the resonator (Fig.2 (b)) is determined from Eq. (2) as well. For a 100 ps Gaussian pulse, which at telecommunication wavelength 1.5 μm ($\omega_0 = 200$ THz) has the spectral width $\Delta\omega = 4.4$ GHz, Eq. (2) yields $\Delta r_{eff} = 0.44$ nm. Thus, a nanometre-shallow bottle resonator can fully confine the 100 ps optical pulses, while opening and closing the resonator requires just a nanometre-high variation of the effective fibre radius.

As opposed to adiabatically slow tuning, which, ideally, is reversible and therefore allows acquiring and releasing a pulse without distortion [6], we do not assume here that the switching process is slow. Instead, we show that it is possible to introduce a fast deformation (refractive index variation) of the bottle resonator without disturbing (i) the optical pulse shape, (ii) the global harmonicity of the potential, and, as a consequence, (iii) the reversibility of the buffering process. To arrive at such an *ideal microscopic optical buffer*, the time-dependent potential $V(z,t)$ is constructed as follows. First, we request that, when fully opened, this potential coincides with a harmonic semi-parabolic potential $V_{op}(z) = \omega_1^2 z^2$, $\omega_1 = k(Rr_0)^{-1/2}$, proportional to the effective radius variation of the bottle resonator $\Delta r_{eff}(z) = -z^2/(2R)$ with axial radius $R$ (Fig. 3(a) and (b), blue curves). Next, we look for the parabolic closed potential centred at point $z = z_0$ in the form $V_{cl}(z) = \omega_2^2(z - z_0)^2 + V_0$ and choose $V_0 = \omega_1^2(1 - \omega_1^2/\omega_2^2)^{-1}z_0^2$ so that these potentials are tangent at their common point $z_c = z_0(1 - \omega_1^2/\omega_2^2)^{-1}$ (Fig. 3(a) and (b), dashed red curves). Finally, the time-dependent buffering potential $V(z,t)$ is constructed as $f(t)V_{op}(z) + (1 - f(t))V_{cl}(z)$, where the switching function $f(t)$ is equal to zero and one for the closed and open potential, respectively. The switching process will not disturb the optical pulse if the characteristic switching time is less than the time it takes the pulse to reside in the left hand side region of $V_{cl}(z)$ near the common point $z_c$ where potentials $V_{op}(z)$ and $V_{cl}(z)$ are approximately equal.

As an example, we numerically investigate a miniature optical buffer having the open and closed configurations defined by the harmonic semi-parabolic and parabolic effective radius variations shown in Fig. 3(a). Buffering of a 100 ps Gaussian pulse is shown in the surface plot of Fig. 3(b). The bottle resonator captures the pulse between the 1st and 2nd ns after launch, holds it over the duration of 4 oscillation cycles and releases with practically no distortion between the 12th and 13th ns. The total delay time is 14 ns, while the time of one cycle is 3.6 ns. Comparison of Fig. 3(a) and (b) shows that, in agreement with Eq. (2), the pulse is fully confined by a parabolic resonator which is as shallow as 1 nm in effective radius variation.

## 4. Microscopic optical buffering in a generalized harmonic non-parabolic potential

In the previous section, we showed that the time-dependent perturbation of an optical pulse can be very small if the pulse is situated in the far right region of the buffer in the process of switching. Here we show that it is possible to design the radius variation profile of a SNAP bottle resonator so that this perturbation is excluded completely along the major part of the buffer.

The parabolic potential is not the only one that possesses the equidistant spectrum and supports the periodic oscillations similar to those shown in Fig. 1(d). We assume here that the potential well is wide enough to ensure a relatively large oscillation time and high enough to ensure the sufficient bandwidth of the pulse. Thus, we are interested in potentials which can be treated semi-classically [19]. We recall that in classical mechanics the potential wells with equidistant spectrum correspond to those having the period of oscillations independent of amplitude. There exist a wide range of such potentials, while a quadratic potential is their simplest representative. Consider a classical particle with energy $E$ and mass $m$ oscillating in a potential well $V(z)$ between turning points $z_1(E)$ and $z_2(E)$. The period of oscillations $T(E)$ is defined by the integral [23]:

$$T(E) = \sqrt{2m} \int_{z_1(E)}^{z_2(E)} \frac{dz}{\sqrt{E - V(z)}} \tag{3}$$

This equation can be also considered as an integral equation (Abel integral equation [3]) which determines potential $V(z)$ for the given dependence of the period on the energy $T(E)$. Analytical solution of this equation allows to express the inverse function $z(V)$ through $T(E)$ [23, 24]. Since $V(z)$ is a potential well, we assume that the function $z(V)$ is two-valued and can be separated into two monotonic branches $z^{(1)}(V)$ and $z^{(2)}(V)$ equal to each other at a common minimum $z_0$, where $V(z_0) = V_0$ and $z^{(1)}(V_0) = z^{(2)}(V_0) = z_0$. Then solution of the integral Eq. (1) yields [23]:

$$z^{(2)}(V) - z^{(1)}(V) = \frac{1}{\pi\sqrt{2m}} \int_{V_0}^{V} \frac{T(E)dE}{\sqrt{V - E}} \tag{4}$$

Here we are searching for a harmonic (but not necessarily quadratic) potential, i.e., the potential having the oscillation period independent of energy, $T(E) \equiv T_0$. In this case, Eq. (4) is simplified and yields the family of harmonic potentials defined by the algebraic equation:

$$z^{(2)}(V) - z^{(1)}(V) = C\left(\sqrt{V - V_0}\right) \tag{5}$$

where $C = (2/m)^{1/2} T_0 / \pi$. In this equation one branch of the potential, e.g., $z^{(1)}(V)$, can be an arbitrary monotonic function, while the other branch, $z^{(2)}(V)$, is expressed through $z^{(1)}(V)$ from Eq. (5). As an example, we construct the buffer potential $V(z,t)$ (or, equivalently, the effective radius variation $\Delta r_{eff}(z,t) = -(r_0 / 2k^2)V(z,t)$) as follows. First, we request that, when fully opened, this potential coincides with a harmonic semi-parabolic potential

$$V_{op}(z) = \omega_1^2 z^2, \tag{6}$$

where $\omega_1 = k(Rr_0)^{-1/2}$ proportional to the effective radius variation of the bottle resonator $\Delta r_{eff}(z) = -z^2 / (2R)$ with axial radius $R$ (Figure 5(a), blue curve). In the process of switching, we request that the left hand side region, $z > z_0$, of potential $V(z,t)$, remains unchanged so that the optical pulse propagating in this region at this time is *not perturbed by switching at all*. This is different from the model considered in main text where the transition between the closed and open potentials deformed the closed potential and therefore slightly perturbed the optical pulse. Next, we request that, after full closing, the new left hand side of the potential, at $z < z_0$, together with the remaining right hand side $z > z_0$ form a closed harmonic potential $V_{cl}(z)$ (Figure 5(a), green dashed curve) where the optical pulse can oscillate without distortion. The analytical expression for $V_{cl}(z)$, is determined from Eq. (5) as:

$$V_{cl}(z) = \begin{cases} \omega_1^2 z^2, & z \geq z_0 \\ \left\{\frac{\omega_1 \omega_2}{\omega_1^2 - \omega_2^2}\left[\omega_1 z - \sqrt{\omega_1^2 z_0^2 + \omega_2^2(z^2 - z_0^2)}\right]\right\}^2 + \omega_1^2 z_0^2, & z < z_0 \end{cases} \tag{7}$$

where $z_0$, $\omega_1$, and $\omega_2$ are free parameters. As requested, potential $V_{cl}(z)$ has the equidistant spectrum in the semi-classical approximation. It is asymmetric and, though continuous everywhere, has a break of the first derivative at $z = z_0$. Finally, the time-dependent buffering potential $V(z,t)$ is constructed as $f(t)V_{op}(z)+(1-f(t))V_{cl}(z)$ as described in the main text. The switching process will not disturb the pulse if the characteristic switching time is less than the time it takes the pulse to reside in the region $z > z_0$.

The buffering process is described as follows. The optical pulse is launched into the open semi-parabolic potential. When the pulse is approaching the right hand side turning point, the semi-parabolic potential is gradually closing and transforming into the asymmetric harmonic potential determined by Eq. (7) (Figure 5(a)). Crucially, this deformation does not affect the right hand side of the potential, which is the *common parabolic part* of both the closed and open harmonic potentials. For this reason, the deformation does not perturb the pulse, which at the time of deformation is situated completely within the right hand side parabolic part of the potential well. For the next period of time, the pulse is oscillating between turning points in the closed potential without distortion. Finally, the inverse time-dependent process transforms the closed parabolic potential into the open semi-parabolic potential to let the pulse out. Again, this process does affect the shape of the optical pulse.

As an example, we numerically investigate a miniature optical buffer having the opened and closed configurations defined by the harmonic semi-parabolic and asymmetric effective radius variations shown in Figure 5(a). Buffering of a 100 ps Gaussian pulse is shown in the surface plot of Figure 5(b). The bottle resonator captures the pulse, holds it over the duration of 6 oscillations and releases with practically no distortion. For the 2 nm high effective radius variation considered, the single oscillation time is 1.7 ns and the whole delay is around 16 ns. The slight discrepancy between the input and output pulses shown in Figure 5(c) can be explained by the fact that the closed potential was determined in the semi-classical approximation, i.e., it is not the exact harmonic potential. However, the remarkably small distortion shown in Figure 5(c) is obviously sufficient for practical applications. In addition, it is expected that the potential profile can be iterated to minimize the distortion further.

## 5. Fabrication precision of SNAP structures

It is instructive to discuss the physical meaning of the dramatically high subangstrom precision, which has been achieved in SNAP technology [18, 25], and its prospective improvement to 0.1 Å, which is required for the realization of the proposed harmonic optical buffer. The value 0.1 Å is an order of magnitude less than the size of an atom. The definition of such a small measurement precision assumes averaging of the actual surface height variation over the surface dimensions much greater the radiation wavelength. Experimentally, we determine the effective radius variation of the fibre using a microfibre taper connected to the power source and optical spectrum analyser as illustrated in Fig. 2(a). The microfibre is translated along the SNAP fibre and the WGM spectrum is measured at sequential points of contacts. In the simplest case of an optical fibre with the slow-varying radius, the radius variation $\Delta r_{eff}(z)$ is determined from the shift a WGM resonance $\Delta\omega(z)$ by equation $\Delta r_{eff}(z)/r_0 = \Delta\omega(z)/\omega_0$ [26, 27]. At characteristic optical frequency of $\omega_0 \sim 200$ THz (corresponding to the telecommunication wavelength 1.6 µm) the width of the resonance for a silica fibre can be as small as $\Delta\omega \sim 10$ MHz (corresponding to Q-factor $\sim 10^7$). With the same OSA resolution of $\Delta\omega \sim 10$ MHz and fibre radius $r_0 \sim 20$ µm, the measurement precision of the fibre radius variation can be as small as $\Delta r \sim \Delta\omega \cdot r_0 / \omega_0 \sim 0.01$ Å.

## 6. Required fabrication precision

Since the absolutely precise harmonic potential cannot be realized experimentally, it is important to understand whether the precision required for satisfactory performance of the buffer is feasible. We verify the effect of deviation from the harmonicity by adding a perturbation (a) inside the closed harmonic potential, which disturbs the process of periodic oscillations and (b) at the entrance of the open harmonic potential, which affects the process of entering and exiting of the resonator. Calculations show that perturbations of the closed harmonic potential, especially those localized near the turning points of the pulse, have a much stronger effect on the distortion of the optical pulse than those of the open semi-parabolic potential. In fact, the pulse speed near the turning points tends to zero, so that the acquired effect of a perturbation here is maximized. Furthermore, the effect of perturbation of the closed potential is multiplied by the number of oscillation cycles. We assume that the perturbations are spatially smooth and choose them in the form of a Gaussian function with FWHM equal to 0.6 mm which corresponds to the characteristic width of the laser beam used for fabrication and tuning of the bottle resonator. Fig. 4 compares the input 100 ps pulse (top, (a)) and the output pulse for the open and closed bottle unperturbed (bottom, (b)); closed bottle perturbed at its right hand side near the turning point (bottom, (c), (d), and (e)); and open bottle perturbed at its left hand side (bottom, (f), (g)). It is seen that, while the 0.1 Å perturbation of the closed bottle introduces minor distortion of the pulse (Fig. 4(c)), the distortion becomes significant for the 0.3 Å perturbation (Fig. 4(d)) and becomes severe for the 0.5 Å perturbation (Fig. 4(e)). The perturbation profile and 2D plot clarifying the pulse evolution for the last case are shown in Figs. 3(c) and (d). In contrast, the perturbation near the exit of the open bottle resonator up to 2 Å, i.e., ~ 10% of the effective radius variation (Fig. 4(f), (g) and Fig. 3(e), (f)) does not cause significant pulse distortion.

## 7. Discussion and summary

Following the original idea of Schrödinger and based on the recent progress in microphotonics, we have introduced and investigated a feasible microscopic optical buffer. It is shown that a few nanometre tuning of the bottle resonator effective radius is sufficient to trap, hold, and release telecommunication optical pulses without distortion over the time period of ten of nanoseconds or longer, while the delay time is limited by the material losses only. The dimensions of this device are determined by the footprint of the SNAP bottle resonator (0.12 $mm^2$ for the model considered, which corresponds to 0.009 $mm^2$ per a nanosecond delay), while each oscillation of the pulse in this resonator delays light by 3.6 ns (compare with a single-cycle untunable semi-parabolic delay line experimentally demonstrated in [18], which had the total delay of 2.6 ns and the same footprint). Remarkably, the parabolic profile of the bottle resonator is not the only profile that allows to hold a light pulse with minimal distortion introduced. There exists a wide family of potential wells in which the period of classical oscillations does not depend on the amplitude [23] and the optical pulse experiences practically no distortion. We have shown that exploiting these generalized potentials allows to optimize the performance of miniature optical buffers more efficiently.

We suggest that the exceptionally high precision of 0.1 Å required for the fabrication of the bottle resonator buffer described above is feasible using the advanced SNAP technology. In fact, the precision of 0.7 Å in effective radius variation was experimentally achieved in [25] by iterations using 0.07 Å increments limited by the resolution of the optical spectrum analyser (OSA) used. Thus, it is expected that the 0.1 Å precision can be achieved by the straightforward improvement of the SNAP fabrication setup. The tunability of the proposed miniature optical buffer can be achieved utilizing the fibre with a highly nonlinear, electrostrictive, or piezoelectric core [28-31]. The local application of a laser field or electric potential to the nonlinear material positioned inside the fibre allows to deform the fibre, tune its effective radius, and, thus, open and close the bottle resonator. For example, tuning has been recently demonstrated for a silica WGM resonator with a silicon core [28, 29]. The wavelength shift as large as 0.4 nm was introduced by a picosecond laser field and attributed to the Kerr nonlinearity of silicon. This shift corresponds to 5 nm of the effective radius variation for a fibre with the 20 μm radius, which is sufficient to enable the buffering process describe above. Since the required spatial distribution of the switching deformation is smooth (Fig. 3(a)), the corresponding intensity variation is feasible. Alternatively, the required nanoscale temporal and spatial variations can be introduced in a fibre segment wholly fabricated of a low loss and highly nonlinear, electrostrictive, and piezoelectric materials (e.g., of silicon [28, 29] or lithium niobate [31]) for which the SNAP technology can potentially be developed. In this case, the power of the switching laser pulse can be enhanced dramatically if the pulse is resonantly coupled into the fibre WGM [32]. The recent demonstration of fully reconfigurable subangstrom-precise SNAP bottle resonator temporary introduced by heating [33] supports the feasibly of the microscopic optical buffer suggest in this paper. Fabrication of highly nonlinear and electrostrictive fibre segments with the required outstanding uniformity is a fruitful and challenging problem to be addressed in the future.

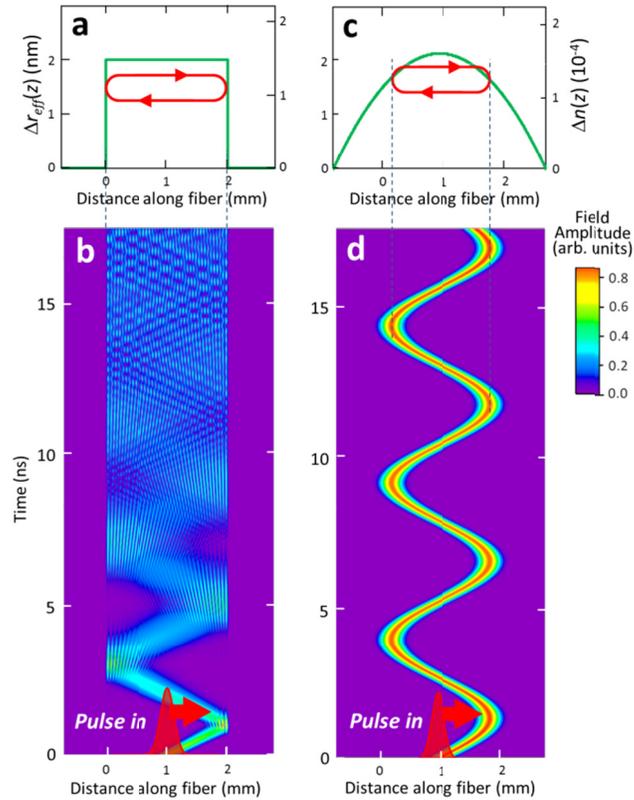

**Figure 1.** Propagation of a 100 ps optical pulse in rectangular and parabolic bottle resonators. (a) Effective radius variation of the rectangular resonator having 2 nm height and 2 mm length. (b) Propagation of a 100 ps optical pulse in this resonator. The surface plot shows the field distribution of the pulse as a function of the coordinate along the fibre and time. It is seen that the pulse is completely corrupted after several nanoseconds of propagation and a few reflections due to the dispersion and self-interference. (c) Effective radius variation of the parabolic resonator having the 345 m radius of curvature. (d) Propagation of a 100 ps pulse in this resonator showing the periodic oscillations of the pulse with no distortion over the oscillation cycles.

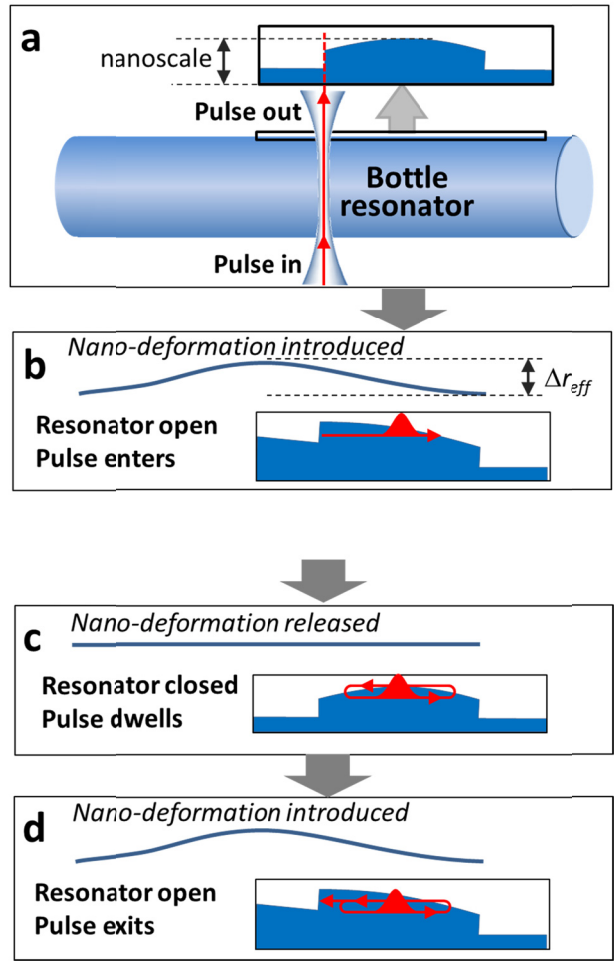

**Figure 2.** Illustration of a SNAP bottle resonator optical buffer. (a) A SNAP bottle resonator created from an optical fibre with nanoscale parabolic radius variation. The resonator is coupled to the transverse input-output waveguide (micrometer-diameter waist of an optical fibre taper). (b) The switching nano-deformation of the effective radius, which is introduced by a pulse of the applied laser or electrical field, transfers the closed parabolic resonator into the open semi-parabolic resonator shown in this figure. (c) After the deformation shown in Figure (b) is released, the bottle resonator restores its original parabolic shape with the optical pulse oscillating inside it. (d) Finally, the same as in Figure (b) nano-deformation is introduced again and the pulse is released back into the input-output waveguide. The red curves illustrate the propagation of an optical pulse in each of the configuration considered.

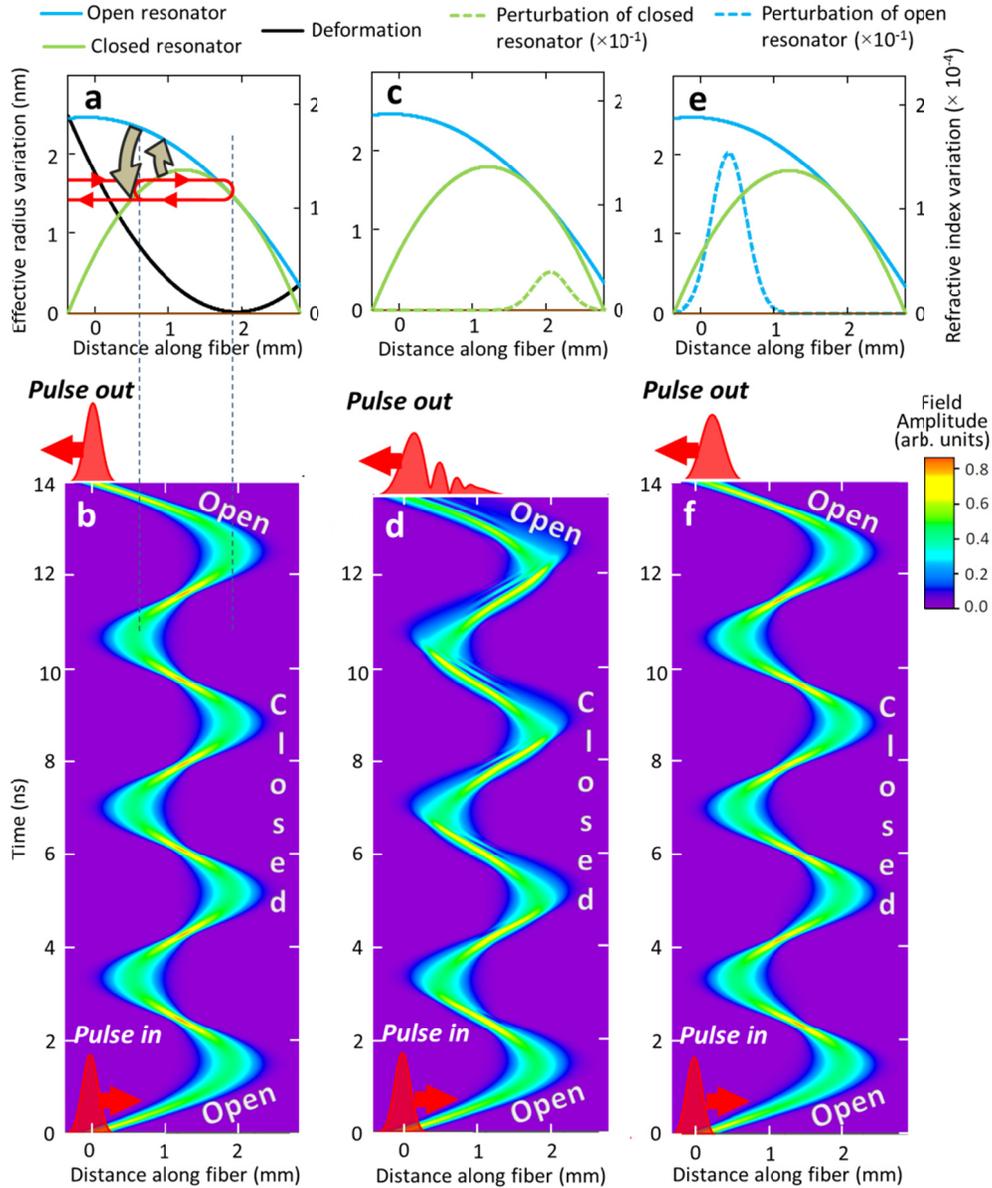

**Figure 3.** Performance of a parabolic SNAP bottle resonator optical buffer. (a) The effective radius variation (left vertical axis) and equivalent refractive index variation (right vertical axis) of the open semi-parabolic resonator (blue curve) and closed parabolic resonator (green curve). The nano-deformation (black curve), equal to the difference of these curves, is gradually introduced and released during a sub-nanosecond time period. (b) The surface plot shows the distribution of the field of the optical pulse, which is captured, held, and released by the buffer, as a function of the coordinate along the bottle resonator and time. The output pulse shown at the top of the figure (and also shown in Fig. 4(b)) exhibits the negligible distortion compared to the input pulse at the bottom. (c) In this figure, the closed parabolic resonator shown in Figure (a) is perturbed by a Gaussian deformation with the height of 0.5 Å (dashed green curve). (d) The surface plot in this figure shows that the optical buffer with such perturbed radius variation exhibits significant distortion of the pulse over time (the output pulse for this case is also shown in Fig. 4(e)). (e) This figure shows a much greater 2 Å Gaussian perturbation with the same width introduced at the entrance of the open semi-parabolic resonator (dashed blue curve). (f) The evolution of the pulse in the optical buffer with such perturbation exhibits the tolerable distortion of the pulse (shown at the top of (f) and also in Fig. 4(g)).

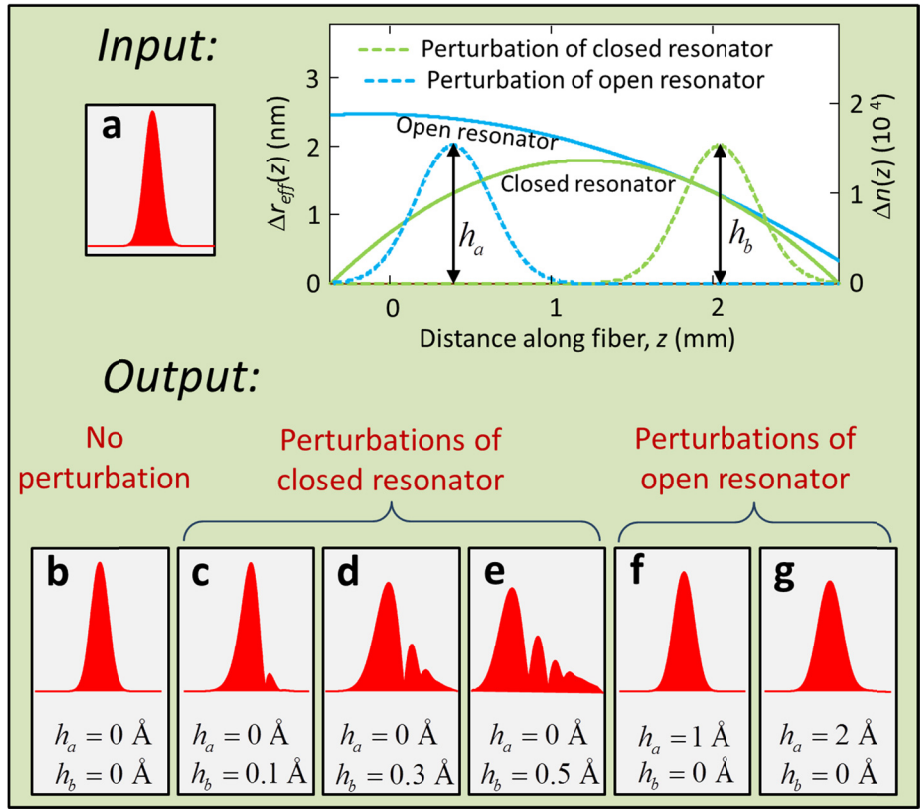

**Figure 4.** Comparison of the input 100 ps pulse (a) and output pulses corresponding to different perturbations of the optical buffer (b)-(g). The profiles of 0.6 mm FWHM Gaussian perturbations of the closed parabolic resonator (green dashed curve with height $h_a$) and open semi-parabolic resonator (blue dashed curve with height $h_b$) which are positioned as shown in the inset. (b) The output pulse for the unperturbed buffer. (c), (d), and (e) The output pulses for the closed resonator perturbations with $h_a = 0.1$, 0.3, and 0.5 Å, respectively. (f) and (e) The output pulses for the open resonator perturbations $h_b = 1$ and 2 Å, respectively.

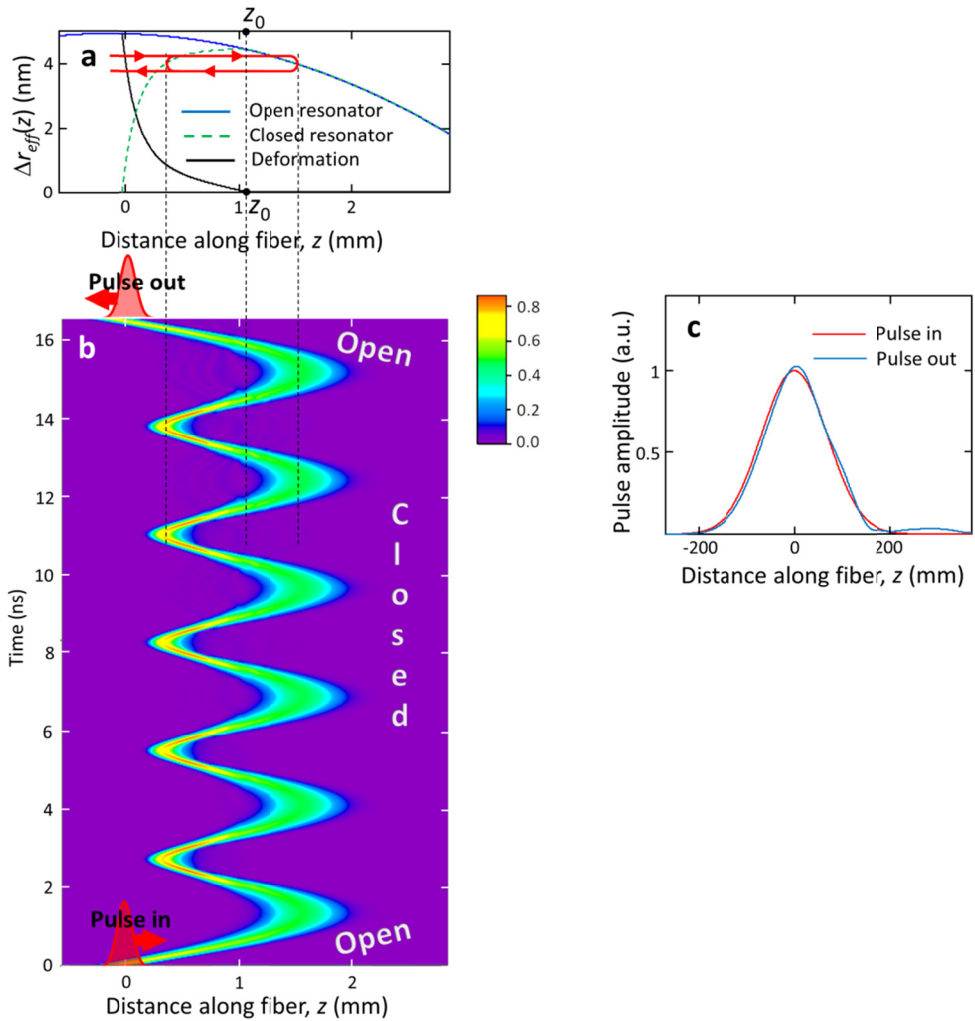

**Figure 5.** Performance of a SNAP bottle resonator optical buffer constructed of an asymmetric semi-classical harmonic non-parabolic potential. (a) The effective radius variation of the open semi-parabolic resonator (blue curve) and closed parabolic resonator (green dashed curve). The nano-deformation (black curve), equal to the difference of these curve, is gradually introduced and released during a sub-nanosecond time period and is equal to zero at $z > z_0$. (b) The surface plot shows the distribution of the field of the optical pulse, which is captured, held, and released by the buffer, as a function of the coordinate along the bottle resonator and time. The output pulse shown at the top of the figure exhibits the negligible distortion compared to the input pulse at the bottom. (c) Comparison of the input and output pulse profiles.